\documentclass[]{aa}  

\usepackage{natbib}

\usepackage{graphicx}
\begin{document}
\title{
Escape-limited Model of Cosmic-ray Acceleration Revisited
}


\titlerunning{Escape-limited Model of Cosmic-ray Acceleration Revisited}
\authorrunning{Ohira et al.}

\author{
Yutaka Ohira
\inst{1},
Kohta Murase
\inst{2},
\and
Ryo Yamazaki
\inst{3}
}

\institute{
Department of Earth and Space Science, Osaka University, 
Toyonaka 560-0043, Japan\\
\email{yutaka@vega.ess.sci.osaka-u.ac.jp}
\and
Yukawa Institute for Theoretical Physics, Kyoto University,
Kyoto 606-8502, Japan\\
\email{kmurase@yukawa.kyoto-u.ac.jp}
\and
Department of Physical Science, Hiroshima University,
Higashi-Hiroshima 739-8526, Japan\\
\email{ryo@theo.phys.sci.hiroshima-u.ac.jp}
}

\date{}


\abstract
{
The spectrum of cosmic rays (CRs) is affected by
their escape from an acceleration site.
This may have been observed not only in
the gamma-ray spectrum of young supernova remnants (SNRs) such as RX~J1713.7$-$3946,
but also in the spectrum of CRs showering on the Earth.
} 
{The escape-limited model of cosmic-ray acceleration is studied in general.
We discuss the spectrum of CRs running away from the acceleration site.
The model may also constrain the spectral index at the acceleration site and the ansatz with respect to the unknown injection process into the particle acceleration.
}
{
Analytical derivations. 
We apply our model to CR acceleration in SNRs and in active galactic nuclei (AGN), which are plausible candidates of Galactic and extragalactic CRs, respectively. 
In particular, for young SNRs, we take account of the shock evolution with cooling of escaping CRs in the Sedov phase.
}
{The spectrum of escaping CRs generally depends on the physical quantities at the acceleration site, such as the spectral index, the evolution of the maximum energy  of CRs and the evolution of {{\bf 
the normalization factor of the spectrum.
}}
It is found that the spectrum of run-away particles can be both softer and harder than that of the acceleration site. 
}
{
The model could explain spectral indices of both Galactic and extragalactic CRs produced by SNRs and AGNs, respectively, suggesting the unified picture of CR acceleration. 
}


\maketitle
\section{Introduction}
\label{sec:intro}
The origin of cosmic rays (CRs) has been one of the long-standing problems. 
The number spectrum of nuclear CRs observed at the Earth, 
$\mathcal{N} (E) \propto E^{-s}$, shows a break at the ``knee'' energy ($\approx10^{15.5}$~eV), below which the spectral index is about $s \approx -2.7$ \citep{cronin1999}.
Because of the energy-dependent propagation of CRs, the spectral shape at the source is different from that observed at the Earth. Taking into account the propagation effect, the source spectral index has been well constrained as $s \approx 2.2-2.4$ in various models \citep[e.g.,][]{strong98,putze2009}.
This value of $s$ has been also inferred in order to explain the 
Galactic diffuse gamma-ray emission \citep[e.g.,][]{strong00}.
This fact may give us valuable insights on the acceleration mechanism of CRs.

Mechanisms of CR acceleration have also been studied for a long time and the most plausible process is a diffusive shock acceleration (DSA) \citep{krymsky77,axford77,bell78,blandford78}.
Very high-energy gamma-ray observations have revealed that the existence of high-energy particles at the shock of young supernova remnants (SNRs), which supports the DSA mechanism as well as the paradigm that the Galactic CRs are produced by young SNRs \citep[e.g.,][]{enomoto02, aharonian04, aharonian05c, katagiri05}.
Recent progress of the theory of DSA has revealed that the back-reactions of accelerated CRs are important if a large number of nuclear particles are accelerated \citep{drury81,malkov01}.
There are several observational facts which are consistent with the predictions of such nonlinear model \citep{vink03, bamba03b, bamba05a, bamba05b, warren2005, uchiyama07, helder2009}.
The model predicts, however, the harder spectrum of accelerated particles at the shock than $f (p) \propto p^{-4}$ (corresponding to $s=2$) where $p$ is the momentum of CRs, in particular, near the knee energy \citep{malkov97, berezhko99, kang01}.
This fact apparently contradicts with the source spectral index of 
$s \approx {2.2-2.4}$ inferred from the CR spectrum at the Earth.
Even in the test-particle limit of DSA, such a soft source spectrum requires a shock with the small Mach number ($M \la10$), 
which is unexpected for young SNRs.

There are several models of DSA, depending on the boundary conditions imposed.
Different models predict different spectra of CRs dispersed from the shock region.
So far, the age-limited acceleration has been frequently considered as a representative case (\S~\ref{sec:agelimit}).
In this model, all the particles are stored around the shock while accelerated.
When the confinement becomes inefficient, all the particles run away from the region at a time.
Then, the source spectrum of CRs which has just escaped from the acceleration region is expected to be the same as that at the shock front. 
Therefore, this model predicts that the source spectrum is the same as 
that of accelerated particles, which is typically harder than the observed one.
In this paper, we consider an alternative model, the escape-limited
acceleration, to explain the observed CR spectrum at the Earth
(\S~\ref{sec:escape}).
This model is preferable to the age-limited acceleration when we consider observational results for young SNR RX~J1713.7$-$3946, of which TeV $\gamma$-ray emission is more precisely measured than any others (\S~\ref{sec:rxj1713}).

The nature of CRs with energies much higher than the knee energy is also still uncertain.
While CRs below the second knee ($\sim {10}^{17.5}$~eV) may be
 Galactic origin, the highest energy CRs above $\sim{10}^{18.5}$~eV are believed to be extragalactic.
Possible candidates are active galactic nuclei (AGNs) \citep[e.g.,][]{BS87,Tak90,RB93,PMM09}, gamma-ray bursts \citep[][]{Wax95,Vie95,MINN06}, magnetars \citep[][]{Aro03,MMZ09}
and clusters of galaxies \citep[][]{KRB97,ISMA07}.
The intermediate energy range from $\sim {10}^{17.5}$~eV to $\sim{10}^{18.5}$~eV is more uncertain. 
Both the Galactic and extragalactic origins are possible and it may just a transition between the two. 
As the extragalactic origin, AGNs \citep{berezinsky06}, clusters of galaxies \citep{MIN08} and hypernovae \citep{WRMD07} have been proposed so far.

Among these possibilities, AGN is one of the most plausible candidates for accelerators of high-energy CRs, because it can explain the UHECR spectrum above $\sim 10^{17.5}$~eV assuming the proton composition. 
In such a proton dip model, the source spectrum of UHECRs is 
$\mathcal{N} \propto E^{-s}$ with $s=2.4$--2.7, depending on models 
of the source evolution \citep{berezinsky06}. 
The required source spectral index of $s=2.4$--2.7 can be explained by several possibilities. 
First, it can be attributed to the acceleration mechanism itself. 
One can consider non-Fermi acceleration mechanisms \citep{berezinsky06} or the two-step diffusive shock acceleration in two different shocks \citep{Alo+07}. 
Second, the index can be attributed to a superposition of many AGNs with different maximum energies, and one can suppose that AGNs with different luminosities may have different maximum energies \citep{KS06}. 
Recently, \citet{berezhko2008}
 proposed another possibility under the cocoon shock model. 
In this cocoon shock scenario, different maximum energies can be 
interpreted as maximum energies of escaping particles at different ages of AGN jets. 
Although it is very uncertain whether efficient CR acceleration occurs
there, this scenario would also be one of the possibilities to be
investigated in detail.

The organization of the paper is as follows.
After the brief introduction of the age-limited model of the CR 
acceleration (\S~\ref{sec:agelimit}), we study the escape-limited model in \S~\ref{sec:escape}.
For a simple understanding, the general argument in a stationary, test-particle approximation is given in \S~\ref{app:testparticle}.
Then, we derive the formulae of the maximum energy of accelerated particles in \S~\ref{sec:pmax_escape}, and of the spectrum of escaping particles in \S~\ref{sec:spect_escape}.
We consider the applications to young SNR and AGN in \S~\ref{sec:snr} and \S~\ref{sec:AGN}, respectively.
Section~\ref{sec:discussion} is devoted to a discussion.

\section{Maximum Attainable Momentum in the Age-limited Acceleration }
\label{sec:agelimit}
For comparison with the escape-limited acceleration, we briefly summarize the case of the age-limited acceleration.
In this case, the maximum momentum of accelerated particles, $p_\mathrm{m}^\mathrm{(age)}$, is determined by $t_\mathrm{acc}=t_\mathrm{age}$, where $t_\mathrm{age}$ and $t_\mathrm{acc}$ are the age of the shock and the acceleration time scale, respectively.
When we consider DSA, $t_\mathrm{acc}$ is given by \citep{drury83}
\begin{equation}
t_{\rm acc} = \frac{3}{u_1-u_2}
\left(\frac{D_1(p)}{u_1}+\frac{D_2(p)}{u_2}\right)~~,
\end{equation}
where $D(p)$ and $u$ are the diffusion coefficient as a function of the momentum of accelerated particles and the velocity of the background fluid, respectively. Subscripts 1 and 2 represent upstream and downstream regions, respectively.
For simplicity, we assume the Bohm-type diffusion, i.e.,
\begin{equation}
D_1(p)=D_2(p)= \frac{\eta_\mathrm{g}m_\mathrm{p}c^3}{3eB}\left(\frac{p}{m_\mathrm{p}c}\right)~~,
\label{eq:diff_age}
\end{equation}
where $B$, $\eta_\mathrm{g}$ and $m_\mathrm{p}$ are the magnetic field strength, the gyro-factor and the proton mass, respectively. 
Taking into account that the fluid velocities are related to the shock velocity, $u_\mathrm{sh}$, as $u_1 = \sigma u_2=u_\mathrm{sh}$, where $\sigma$ is the compression ratio of the shock, we derive \citep[e.g.,][]{aharonian99}
\begin{equation}
p_\mathrm{m}^\mathrm{(age)} = \frac{\sigma-1}{\sigma(\sigma+1)}\frac{eB}{\eta_\mathrm{g}c^2}u_\mathrm{sh}^2t_\mathrm{age}~~.
\label{eq:pmax_age}
\end{equation}
%

\section{Escape-limited Acceleration}
\label{sec:escape}

In the frame work of DSA, accelerated particles are scattered by the turbulent magnetic field, and go back and forth across the shock front.
Upstream turbulence may be excited by the accelerated particles themselves \citep{bell78}, and the magnetic field strength of such turbulence is theoretically expected to be strong \citep[e.g.,][]{lucek00}.
There are observational evidences suggesting that CRs are responsible for substantial amplification of the ambient magnetic field in the precursors of shock fronts in SNRs, and that such magnetic turbulence well confines the particles around the shock front \citep{vink03, bamba03b, bamba05a, bamba05b, yamazaki04, parizot06, uchiyama07}, leading to the efficient CR acceleration.

The spectrum of accelerated particles is affected by the spatial and spectral structures of the magnetic turbulence through the process in which the particles escape from the shock toward far upstream region.
There are mainly two scenarios of the escape model considered so far; 
one causes the effect on the boundary in the momentum space, 
and the other causes the effect on the spatial boundary.
The former comes from significant decay of the wave amplitude below the wave number $k_{\min}$ of the spectrum of the turbulence \citep{reynolds98,drury09}.
Particles with the Lorentz factor above $\gamma_\mathrm{c}$ satisfying the resonance condition, $\omega-k_{\min}v-\Omega_\mathrm{c}(\gamma_\mathrm{c})=0$, 
where $\Omega_\mathrm{c}(\gamma)=eB/\gamma m_{\mathrm{p}}c$ is the cyclotron frequency, are not confined around the shock front and escape into far upstream region. In this context, the escape flux was calculated previously  \citep[e.g.,][]{ptuskin05,drury09}.
The latter effect has been recently discussed by several authors 
\citep{ptuskin05, reville09, caprioli09}.
The turbulence generation may be connected with the flux of accelerated particles themselves. Hence, in the region far from the shock front, the flux of high-energy particles is small and wave excitation is less significant. 
If the accelerated particles reach the region, they are dispersed into the far upstream region.
Let $\ell$ be the distance from the shock beyond which the amplitude of the upstream turbulence becomes negligible.
Characteristic spatial length of particles penetrating into the upstream 
region is given by $D(p)/u_\mathrm{sh}$.
As long as $D(p)/u_\mathrm{sh}\ll\ell$, the particles are confined 
without the significant escape loss, and they are accelerated to higher energies.
On the other hand, when their momentum increases up to sufficiently high energies satisfying $D(p)/u_\mathrm{sh}\ga \ell$, their acceleration ceases and they escape into the far upstream.
Therefore, the maximum momentum of accelerated particles in this 
scenario is given by the condition $D(p)/u_{\rm sh}\sim\ell$.
In the following of the paper, we consider the escape-limited model, 
where the maximum energy is essentially determined by $D(p)/u_\mathrm{sh} \la \ell$.

\subsection{A simple case of stationary, test-particle approximation}
\label{app:testparticle}
In order to take an essential feature of the escape-limited acceleration, 
we calculate the escape flux and the maximum attainable energy of accelerated particles in the simplest case \citep[see also][]{caprioli09}.
Let us consider the stationary transport equation
\begin{equation}
u(x)\frac{\partial f}{\partial x}=
\frac{\partial}{\partial x}\left[
D(p)\frac{\partial f}{\partial x}\right]
+\frac{p}{3}\frac{\mathrm{d} u}{\mathrm{d} x}\frac{\partial f}{\partial p} +Q~~,
\label{eq:diffusion_convection}
\end{equation}
with the boundary condition,
$f(x=-\ell)=0$, where $x=-\ell$ ($\ell>0$) is the upstream escape boundary.
The fluid velocity $u(x)$ is given by
\begin{equation}
u(x) =
\left\{
\begin{array}
{c@{} l@{}}
u_1 &\ \  (x<0) \\
u_2 &\ \  (x>0) 
\end{array} \right. ~~,
\label{eq:velocity_step}
\end{equation}
where $u_1$ and $u_2$ are constants.
The solution to the transport equation in the test-particle approximation is derived as \citep{caprioli09}
\begin{equation}
f(x,p)=f_0(p)
\frac{\exp\left[\frac{u_1x}{D(p)}\right]-\exp\left[-\frac{u_1\ell}{D(p)}\right]}
{1-\exp\left[-\frac{u_1\ell}{D(p)}\right]}~~,
\end{equation}
where $f_0(p)=f(x=0,p)$ is given by
\begin{equation}
f_0(p)=K\exp\left\{
-q\int_{p_\mathrm{inj}}^p\frac{\mathrm{d}\log p'}
{1-\exp\left[-\frac{u_1\ell}{D(p')}\right]}
\right\}~~,
\label{eq:f0testparticle}
\end{equation}
and $q=3u_1/(u_1-u_2)$.
The escape flux at $x=-\ell$ is
\begin{eqnarray}
\phi(p) &=& 
u_1 \left. f \right|_{x=-\ell}-D(p)
\left. \frac{\partial f}{\partial x} \right|_{x=-\ell} \nonumber \\
&=& \frac{u_1f_0(p)}{1-\exp\left[\frac{u_1\ell}{D(p)}\right]}~~.
\end{eqnarray}

Let us introduce a new variable $y=\ln p$ and a new function $\Phi(y)=\ln\phi(p)$.
Then, we expand $\Phi(y)$ around its maximum value at $y=y_\mathrm{m}$.
To do this, we calculate the first and the second derivatives as
\begin{equation}
\frac{\mathrm{d}\Phi}{\mathrm{d} y} = -\frac{1}{1-\exp\left[-\frac{u_1\ell}{D(y)}\right]}
\left[q-\frac{u_1\ell}{D(y)}\frac{\mathrm{d}\ln D}{\mathrm{d} y}\right]~~,
\end{equation}
\begin{eqnarray}
&&\frac{\mathrm{d}^2\Phi}{\mathrm{d} y^2} = \frac{\frac{u_1\ell}{D(y)}}
{\left\{1-\exp\left[-\frac{u_1\ell}{D(y)}\right]\right\}^2}
\left\{
\left[q-\frac{u_1\ell}{D(y)}\frac{\mathrm{d}\ln D}{\mathrm{d} y}
\right]
\frac{\mathrm{d}\ln D}{\mathrm{d} y} \right.\nonumber\\
&&
\left. \ \ \ \ \ \ 
+ 
\left[1-\exp\left[-\frac{u_1\ell}{D(y)}\right]\right]
\left(\frac{\mathrm{d}\ln D}{\mathrm{d} y}\right)^2-
\frac{\partial^2\ln D}{\partial y^2}
\right\}~.
\end{eqnarray}
In the following, we consider the case of Bohm diffusion, $D(p)=D_0(p/m_\mathrm{p}c)$. 
Then, one can find 
\begin{equation}
p_\mathrm{m} = e^{y_\mathrm{m}} = \frac{u_1\ell m_\mathrm{p}c}{qD_0}~~,
\label{eq:pmax_escape}
\end{equation}
(where ${\rm d} \Phi/{\rm d} y=0$) so that we obtain
\begin{eqnarray}
\Phi(y) &=& \Phi(y_\mathrm{m}) + 
 \frac{1}{2}\frac{\partial^2\Phi}{\partial y^2}(y_\mathrm{m})(y-y_\mathrm{m})^2
 + \cdots~~, \nonumber \\
&=& \ln\left[
\frac{u_1f_0(p_\mathrm{m})}{1-e^{-q}}\right]
-\left(\frac{y-y_\mathrm{m}}{\xi}\right)^2 + \cdots~~,
\end{eqnarray}
where $\xi^2 = 2(1-e^{-q})/q<1$.
Note that the quantity $p_\mathrm{m}$ given by Eq.(\ref{eq:pmax_escape}) also plays the role of maximum momentum of the accelerated particles at the shock ($x=0$) because one can see $f_0(p)\propto p^{-q}$ for $p\ll p_\mathrm{m}$ while $f_0(p)\propto\exp(-p/p_\mathrm{m})$ for $p_\mathrm{m}\ll p$ \citep{caprioli09}.

Finally, going back to the function $\phi(p)$, we obtain
\begin{equation}
\phi(p) =
\frac{u_1f_0(p_\mathrm{m})}{1-e^{-q}}
\exp\left[-\left(\frac{\ln p-\ln p_\mathrm{m}}{\xi}\right)^2
+\cdots\right]~~.
\label{eq:phi2}
\end{equation}
One can clearly see, from Eq.~(\ref{eq:phi2}), that particles with momentum around $p_{\rm m}$ escape the shock region most  efficiently.

\subsection{The maximum energy of accelerated particles}
\label{sec:pmax_escape}

We have seen in \S~\ref{app:testparticle} that in the stationary, test-particle case, the quantity $p_\mathrm{m}$ given by Eq.~(\ref{eq:pmax_escape}) plays the role of maximum momentum of the accelerated particles at the shock.
Taking this fact into our mind, we assume that in the more general escape-limited case, the maximum momentum, $p_{\rm m}^{\rm (esc)}$, is determined by
\begin{equation}
q \frac{D(p_\mathrm{m}^\mathrm{(esc)})}{u_1}=\ell~~.
\label{eq:condition_esc}
\end{equation}
Given that
\begin{equation}
D(p)= \frac{\eta_\mathrm{g}m_\mathrm{p}c^3}{3eB}\left(\frac{p}{m_\mathrm{p}c}\right)~~,
\end{equation}
which is the same as in the age-limited case [Eq.~(\ref{eq:diff_age}) in \S~\ref{sec:agelimit}], we obtain
\begin{equation}
p_\mathrm{m}^\mathrm{(esc)} = 
\frac{3}{q}\frac{eB}{\eta_\mathrm{g}c^2}u_\mathrm{sh} \ell~~.
\label{eq:pmax_esc2}
\end{equation}

Since $q=3\sigma/(\sigma-1)$, we find from Eqs.~(\ref{eq:pmax_age}) and
(\ref{eq:pmax_esc2})
\begin{equation}
\frac{p_\mathrm{m}^\mathrm{(esc)}}{p_\mathrm{m}^\mathrm{(age)}}
=(\sigma+1) \frac{\ell}{u_\mathrm{sh}t_\mathrm{age}}~~.
\label{eq:ratio_pm}
\end{equation}
Hence, as long as $\ell\gg u_\mathrm{sh}t_\mathrm{age}$ which is expected in the early phase of the shock, we don't need the escape-limited acceleration. 
However, if the shock evolves so that $\ell\ll u_\mathrm{sh}t_\mathrm{age}$, we have to consider the effect of the particle escape, and we cannot simply apply the well-known result of age-limited acceleration, Eq.~(\ref{eq:pmax_age}).

\subsection{The spectrum of CRs dispersed from an accelerator}
\label{sec:spect_escape}

In this subsection, we derive the time-integrated spectrum of CR
particles which is dispersed from an accelerator. 
The derivation is essentially identical to that of \citet{ptuskin05}.
However, our argument is simpler and more general,
so the final form of the spectrum (Eqs.~(\ref{eq:fesc3}) and (\ref{eq:fesc5}))
 is more general.
Note that our formalism is applicable not only to DSA but also to arbitrary
acceleration processes.

The proton production rate, $N(p,\chi)\mathrm{d}\chi \mathrm{d} p$, at a certain epoch labeled by a parameter $\chi$, is defined as the number of protons with momentum between $p$ and $p+\mathrm{d}p$ which is produced in the interval between $\chi$ and $\chi+\mathrm{d}\chi$.
Here $\chi$ is the parameter which describes the dynamical evolution of the accelerator --- it can be either simply the age, or the position of the shock front.
It is expected that $N(p,\chi)$ contains the term of exponential cutoff at the momentum $p_\mathrm{m}(\chi)$ which depends on $\chi$ [see, for example, Eq.~(\ref{eq:N})].
The number of protons with momentum between $p$ and $p+\mathrm{d}p$ that is escaping from the accelerator at the epoch between $\chi$ and $d\chi$ is denoted by $\phi(p,\chi)\mathrm{d}\chi \mathrm{d} p$, and we assume
\begin{equation}
4 \pi p^2 \phi(p,\chi) \mathrm{d}\chi \mathrm{d} p
\propto N(p_\mathrm{m}(\chi),\chi)G(p,\chi)\mathrm{d}\chi \mathrm{d} p~~,
\label{eq:phi}
\end{equation}
where
\begin{equation}
G(p,\chi)=
\exp\left[-\left(\frac{\ln p-\ln p_\mathrm{m}(\chi)}{\xi}\right)^2\right]~~,
\label{eq:W}
\end{equation}
and $\xi<1$. This is because we expect that particles with momentum around $p_{\rm m}(\chi)$ is most efficiently escaping from the source.
Indeed, as shown in \S~\ref{app:testparticle}, Eqs.~(\ref{eq:phi}) and (\ref{eq:W}) are good approximation in the test-particle, stationary case.
Here, we simply expect these assumptions 
are also correct in general.

The time-integrated spectrum of protons which have escaped at the source, $N_{\rm esc}(p)$, is obtained by
\begin{equation}
N_\mathrm{esc}(p)
=4 \pi p^2 \int \phi(p,\chi)\mathrm{d}\chi~~.
\label{eq:f_esc}
\end{equation}
In order to derive a simple analytical form, we approximate Eq.~(\ref{eq:W}) as
\begin{equation}
G(p,\chi)\approx\sqrt{\pi}\xi\cdot
\delta(\ln p-\ln p_\mathrm{m}(\chi)) ~~,
\label{eq:W2}
\end{equation}
If we use a general mathematical formula for $\delta$-functions for an arbitrary function $g(\chi)$:
\begin{equation}
\delta(g(\chi)) =
\frac{\delta(\chi-\chi_0)}
{\left[\frac{\mathrm{d} g}{\mathrm{d}\chi}\right]_{\chi=\chi_0}}~~,
\end{equation}
where $g(\chi_0)=0$, then Eq.~(\ref{eq:W2}) is rewritten as
\begin{equation}
G(p,\chi)\propto 
\frac{p_\mathrm{m}(\chi)}
{\left[\frac{\mathrm{d} p_\mathrm{m}}{\mathrm{d}\chi}\right]_{\chi=p_\mathrm{m}^{-1}(p)}}\delta(\chi-p_\mathrm{m}^{-1}(p)) ~~,
\label{eq:W3}
\end{equation}
where $p_\mathrm{m}^{-1}(p)$ is the inverse function of $p_\mathrm{m}(\chi)$, that is, $p_\mathrm{m}(p_\mathrm{m}^{-1}(p))=p$ and $p_\mathrm{m}^{-1}(p_\mathrm{m}(\chi))=\chi$.
Using Eqs.~(\ref{eq:phi}) and (\ref{eq:W3}), we calculate $N_\mathrm{esc}(p)$ as
\begin{equation}
N_\mathrm{esc}(p) \propto
\frac{p N(p,p_\mathrm{m}^{-1}(p))}{p_{\rm m}^{-1}(p)
\left[\frac{d p_\mathrm{m}}{\mathrm{d}\chi}\right]_{\chi=p_\mathrm{m}^{-1}(p)}}~~.
\label{eq:fesc4}
\end{equation}
This is the most general analytical formula of the spectrum of protons dispersed from an accelerator.

In the following of the paper, the form of $N(p,\chi)$ is assumed to have
\begin{eqnarray}
N(p,\chi)\mathrm{d}\chi \mathrm{d} p&=&K(\chi)\left(\frac{p}{m_\mathrm{p}c}\right)^{-s} \nonumber \\
&&\times\exp\left[-\left(\frac{p}{p_\mathrm{m}(\chi)}\right)\right]
\mathrm{d}\log\chi\,\mathrm{d} p~~, \nonumber \\
\label{eq:N}
\end{eqnarray}
which is a  power-law with the index $s$ and the exponential cut-off at
$p_\mathrm{m}(\chi)$.
Substituting Eq.~(\ref{eq:N}) into Eq.~(\ref{eq:fesc4}), we obtain
\begin{equation}
N_\mathrm{esc}(p)\propto
\frac{p^{1-s}K\left(p_\mathrm{m}^{-1}(p)\right)}
{p_\mathrm{m}^{-1}(p)
\left[\frac{\mathrm{d} p_\mathrm{m}}{\mathrm{d}\chi}\right]_{\chi=p_\mathrm{m}^{-1}(p)}}~~.
\label{eq:fesc2}
\end{equation}
In particular, if $K(\chi)$ and $p_\mathrm{m}(\chi)$ are written by the
power-law forms, such as $K(\chi)\propto\chi^\beta$ and $p_\mathrm{m}(\chi)\propto\chi^{-\alpha}$, then, $p_\mathrm{m}^{-1}(p)\propto p^{-1/\alpha}$, so that Eq.~(\ref{eq:fesc2}) becomes
\begin{equation}
N_\mathrm{esc}(p) \propto p^{-s_\mathrm{esc}}~~,
\label{eq:fesc3}
\end{equation}
where 
\begin{equation}
s_\mathrm{esc} = s + \frac{\beta}{\alpha}~~.
\label{eq:fesc5}
\end{equation}
This is the simplest form of the spectrum of CRs which are dispersed
from the acceleration region.

Generally speaking, in order to  obtain the energy spectrum of
accelerated particles,  time-dependent kinetic equation should be
solved.
Instead,  we have assumed that at an arbitrary epoch, the spectral form is given by Eq.~(\ref{eq:N}). 
This assumption is justified if the spectrum  at the given epoch 
is dominated by those which are being accelerated at that time,
in other words, if the  particle spectrum  does not so much depend on the past acceleration history.
For example, in the case of the spherical expansion, accelerated particles suffer adiabatic expansion after they are transported downstream of the shock and lose their energy \citep[e.g.,][]{yamazaki06}, so that the contribution of the previously accelerated particles is negligible.
Strictly speaking, even if we consider the energy loss via adiabatic expansion, the energy spectrum of accelerated particles does depend on the past acceleration history in some cases.
When we use  the shock radius $R_\mathrm{sh}$ as $\chi$, the final form of Eqs.~(\ref{eq:fesc3}) and (\ref{eq:fesc5}) is correct as long as 
$\beta>\left(s-1\right)\left(\sigma^{-1}-1\right)$ (see Appendix), which is satisfied in the cases considered in \S~\ref{sec:snr} and \S~\ref{sec:AGN}.
Otherwise, the form of Eq.~(\ref{eq:N}) is no longer a good approximation, and the final form of $s_\mathrm{esc}$ is different from Eq.~(\ref{eq:fesc5}) (see Appendix).

\section{Application to Young Supernova Remnants}
\label{sec:snr}
\subsection{Inconsistency of age-limited acceleration with observed results of RX~J1713.7$-$3946}
\label{sec:rxj1713}
RX~J1713.7$-$3946 is a representative SNR from which bright TeV $\gamma$-rays have been detected.
The H.E.S.S. experiment measured the TeV spectrum and claimed that its shape was better explained by the hadronic model \citep{hess06,hess07}.
Furthermore, evidences of amplified magnetic field ($B\ga0.1$~mG) are derived from the width of synchrotron X-ray filaments \citep[Parizot et al. 2006; see also][]{vink03,bamba03b,bamba05a,bamba05b} and from time variation of synchrotron X-ray hot spots \citep{uchiyama07}.
These facts also support the hadronic origin of TeV $\gamma$-rays,
because the leptonic, one-zone emission model \citep[e.g.,][]{aharonian99} cannot explain the TeV-to-X-ray flux ratio.
Hence it is natural to assume that the TeV $\gamma$-rays are produced by the hadronic process although there are several arguments against this interpretation \citep{butt08,katz08,plaga07}.

In the age-limited case, Eq.~(\ref{eq:pmax_age}) reads
\begin{equation}
cp_\mathrm{m}^\mathrm{(age)}=8\times10^3
\frac{B_\mathrm{mG}}{\eta_\mathrm{g}}
\left(\frac{u_\mathrm{sh}}{4000~\mathrm{km}~\mathrm{s}^{-1}}\right)^2
\left(\frac{t_\mathrm{age}}{10^3\mathrm{yr}}\right)\mathrm{TeV}~~,
\label{eq:Emax_p_age_snr}
\end{equation}
where we adopt $\sigma=4$, and $B_\mathrm{mG}$ is the magnetic field strength in units of mG.

H.E.S.S. observation has revealed that the cutoff energy of TeV $\gamma$-ray spectrum is low \citep{hess06,hess07}, so that in the one-zone hadronic scenario the maximum energy of protons, $E_{\max,p}=cp_\mathrm{m}^\mathrm{(age)}$ is estimated as 30--100~TeV \citep{villante07}.
If $E_{\max,p}<100$~TeV and $B\approx1$~mG, then Eq.~(\ref{eq:Emax_p_age_snr}) tells us $\eta_\mathrm{g}\ga80$, implying far from the ``Bohm limit'' ($\eta_\mathrm{g}\approx1$) which is inferred from the X-ray observation \citep{parizot06,yamazaki04} or expected theoretically \citep{lucek00, bell04, reville07, giacalone07, inoue09, ohira09b,ohira09c}.
This statement is recast if we involve recent results of X-ray observations.
The precise X-ray spectrum of RX~J1713.7$-$3946 is revealed, 
which gives $u_\mathrm{sh}=3.3\times10^8\eta_\mathrm{g}^{1/2}$~cm~s$^{-1}$ \citep{tanaka08}.
Then, Eq.~(\ref{eq:Emax_p_age_snr}) can be rewritten as \citep{yamazaki09}
\begin{equation}
cp_\mathrm{m}^\mathrm{(age)}=5\times10^3
B_\mathrm{mG}
\left(\frac{t_\mathrm{age}}{10^3\mathrm{yr}}\right)\mathrm{TeV}~~.
\end{equation}
Hence, in order to obtain $cp_\mathrm{m}^\mathrm{(age)}<100$~TeV, we need $B\la20~\mu$G  in the context of the hadronic scenario of TeV $\gamma$-rays, which contradicts current estimates of the magnetic field.

A possible solution is to consider the escape-limited acceleration.
One can find from Eq.~(\ref{eq:ratio_pm}) that if we take $\ell\sim10^{-2}u_\mathrm{sh}t_\mathrm{age}\sim0.04$~pc, the maximum energy becomes
\begin{eqnarray}
E_{\max,p} &=&
\min\left\{ cp_\mathrm{m}^\mathrm{(age)}, cp_\mathrm{m}^\mathrm{(esc)} \right\} \nonumber \\
&=& cp_\mathrm{m}^\mathrm{(esc)} \nonumber\\
&\sim& 100~\mathrm{TeV}~~,
\end{eqnarray}
which is consistent with the observed gamma-ray spectrum.
In the following, we consider the model of escape-limited acceleration under simple assumptions, estimating the evolution of the number density and the maximum momentum of accelerated particles so as to discuss the spectral index, $s_\mathrm{esc}$, of $N_\mathrm{esc}(p)$.

\subsection{Evolution of $p_{\rm m}$}
\label{sec:SNRpmax}

Time evolution of the maximum momentum of accelerated particles, 
$p_\mathrm{m}$, has been so far discussed in many contexts \citep[e.g.,][]{ptuskin03}.
One way to estimate $p_\mathrm{m}$ is to use Eq.~(\ref{eq:pmax_esc2}).
In this approach, a key parameter is the magnetic field, which is likely amplified around the shock front \citep{vink03, bamba03b, bamba05a, bamba05b, yamazaki04, uchiyama07} and may depend on various physical quantities such as the shock velocity, the ambient density, and so on. 
At present, the evolution of the magnetic field is not well understood 
despite many works \citep[e.g.,][]{niemiec08, requelme09, ohira09a,LM09}.
In addition, the evolution of another parameter $\eta_\mathrm{g}$ is also unknown.
These facts prevents us from predicting $p_\mathrm{m}$ rigorously.

Here we adopt a different phenomenological approach based on the assumption that young SNRs  are responsible for observed CRs below the knee \citep{gabici09}.
The maximum energy $cp_\mathrm{m}$ is expected to increase up to the knee energy
 ($10^{15.5}$ eV) until the end of the free expansion phase, 
$t_\mathrm{Sedov}$, and decreases from that epoch.
As seen in \S~\ref{sec:rxj1713}, $p_\mathrm{m}$ is limited by the escape 
at $t>t_\mathrm{Sedov}$, that is $p_\mathrm{m}=p_\mathrm{m}^\mathrm{(esc)}$.
Then, to reproduce the observed CR spectrum from $\sim$~GeV to the knee, we may assume a functional form of
\begin{equation}
p_\mathrm{m}^\mathrm{(esc)}(t)=p_\mathrm{knee}
\left(\frac{t}{t_\mathrm{Sedov}}\right)^{-6.5/a}~~,
\end{equation}
where $cp_\mathrm{knee}=10^{15.5}~\mathrm{eV}$, so that $cp_\mathrm{m}^\mathrm{(esc)}=1~\mathrm{GeV}$ at $t=10^at_\mathrm{Sedov}$.
For later convenience, we change variables from $t$ to the SNR radius $R_\mathrm{sh}$ in order to take $\chi=R_\mathrm{sh}$, where $\chi$ is a variable appeared in \S~\ref{sec:spect_escape}.
We further assume the dynamics of $R_\mathrm{sh}$ as
\begin{equation}
R_\mathrm{sh} = R_\mathrm{Sedov}
\left(\frac{t}{t_\mathrm{Sedov}}\right)^b~~,
\end{equation}
where $R_\mathrm{Sedov}$ is the shock radius at $t=t_\mathrm{Sedov}$. Then, we obtain
\begin{equation}
p_\mathrm{m}^\mathrm{(esc)}(R_\mathrm{sh}) = 
p_\mathrm{knee} \left(\frac{R_\mathrm{sh}}{R_\mathrm{Sedov}}\right)^{-6.5/ab}~~,
\label{eq:pm}
\end{equation}
so that we have $\alpha = 6.5/ab$ (see the last paragraph of \S~\ref{sec:spect_escape}).
If we adopt $a\approx2.5$ \textit{ad hoc}, in addition to $b\approx2/5$ that is expected in the Sedov phase, then we find $\alpha\approx6.5$ as a phenomenologically required value in the escape-limited model. 
Note that, even if the SNR dynamics is modified by the CR escape, 
$b$ is almost the same as one in the adiabatic case and its effect on 
$\alpha$ is expected to be so small that our conclusion is not affected qualitatively.

\subsection{Dynamics of SNR shock waves}
\label{sec:SNRshock}

In this subsection, we consider the dynamics of the SNR shock in order to estimate the evolution of the normalization factor of the spectrum $K(R_\mathrm{sh})$.
A simple treatment of the dynamics of the SNR shock from the free
expansion to the adiabatic expansion (Sedov) phase has been given by several authors \citep{ostriker88,drury89,bisnovatyi95}. 
Here we extend their method taking account of the cooling by CR escape. The total mass of the SNR shock shell is calculated as
\begin{equation}
M(R_\mathrm{sh}) = M_\mathrm{ej} + 4\pi \int_0^{R_\mathrm{sh}} 
 \mathrm{d} r r^2 \rho_\mathrm{am}(r)~~,
\label{eq:mass}
\end{equation}
where $M_\mathrm{ej}$ and $\rho_\mathrm{am}(r)$ are the ejecta mass and the density of ambient gas, respectively. 
The equation of motion of the thin shell is given by \citep{ostriker88, bisnovatyi95}
\begin{equation}
\frac{\mathrm{d}(Mu)}{\mathrm{d}t} = 4\pi R_\mathrm{sh}^2 (P_\mathrm{in}-P_\mathrm{am})~~,
\label{eq:eom}
\end{equation}
where the gas velocity $u$ is related to the shock velocity $u_\mathrm{sh}$ and the adiabatic index $\gamma_\mathrm{ad}$ as $u=2u_\mathrm{sh}/(\gamma_\mathrm{ad}+1)$. 
Quantities $P_\mathrm{in}$ and $P_\mathrm{am}$ are the pressures of the post-shock gas and the ambient gas, respectively. 
For strong shocks, one can neglect $P_\mathrm{am}$. 
The explosion energy ${\cal E}={\cal E}_\mathrm{th}+Mu^2/2 +Q_\mathrm{esc}$ consists of the internal energy ${\cal E}_\mathrm{th}=4\pi R_\mathrm{sh}^3P_\mathrm{in}/(3(\gamma_\mathrm{ad}-1))$, the kinetic energy, and the energy $Q_\mathrm{esc}$ which is carried by the escaping CRs. 
Then Eq.~(\ref{eq:eom}) is rewritten as
\begin{equation}
\frac{\mathrm{d}(Mu)}{\mathrm{d}t} +
\frac{3(\gamma_\mathrm{ad}-1)}{2R_\mathrm{sh}}Mu^2
= \frac{3(\gamma_\mathrm{ad}-1)}{R_\mathrm{sh}} \left\{{\cal E} 
 - Q_\mathrm{esc}(R_\mathrm{sh}) \right\}~~.
\label{eq:eom2}
\end{equation}
Since $\mathrm{d} t=2\mathrm{d} R_\mathrm{sh}/(\gamma_\mathrm{ad}+1)u$, the left-hand side of Eq.~(\ref{eq:eom2}) becomes 
\begin{equation}
\frac{\gamma_\mathrm{ad}+1}{4M}R_\mathrm{sh}^{-\omega}
\frac{\mathrm{d}}{\mathrm{d} R_\mathrm{sh}}\left[
R_\mathrm{sh}^\omega(Mu)^2\right]~~,
\end{equation}
where $\omega=6(\gamma_\mathrm{ad}-1)/(\gamma_\mathrm{ad}+1)$.
Hence we obtain
\begin{eqnarray}
u_\mathrm{sh}(R_\mathrm{sh})&=& \frac{\gamma_\mathrm{ad}+1}{2} \left[ \frac{2\omega}{M^2(R_\mathrm{sh})R_\mathrm{sh}^{\omega}} \right. \nonumber \\
&&\times \left. \int_0^{R_\mathrm{sh}} \mathrm{d} r ({\cal E} -Q_\mathrm{esc}(r)) 
r^{\omega -1} M(r) \right]^{1/2}~~.
\label{eq:ush}
\end{eqnarray}
Once $Q_\mathrm{esc}(r)$ and $\rho_\mathrm{am}(r)$ are given, 
we can integrate Eqs.~(\ref{eq:mass}) and (\ref{eq:ush}) to derive the dynamics of the SNR, $u_\mathrm{sh}(R_\mathrm{sh})$.
We mainly consider the evolution of $u_\mathrm{sh}$ in the Sedov phase.
Hence, in the following, we simply assume that $\rho_\mathrm{am}$ is constant with $r$ because even in the case of the core-collapse supernova, the wind region is about a few pc and has been already passed by the SNR shock until the beginning of the Sedov phase.
The energy carried by escaping particles $Q_\mathrm{esc}$ is written as
\begin{equation}
Q_\mathrm{esc}(R_\mathrm{sh}) = \Theta (R_\mathrm{sh}-R_\mathrm{Sedov}) 
\int_{p_\mathrm{m}(R_\mathrm{sh})}^{p_\mathrm{knee}} cp N_\mathrm{esc}(p) \mathrm{d} p~~, 
\label{eq:Qesc}
\end{equation}
where $\Theta(x)$ is the Heaviside step function, i.e., $\Theta(x)=1$
for $x>0$ while $\Theta(x)=0$ for $x<0$. 
Substituting Eqs.~(\ref{eq:fesc3}), (\ref{eq:fesc5}), and (\ref{eq:pm}) into Eq.~(\ref{eq:Qesc}), we obtain 
\begin{eqnarray}
Q_\mathrm{esc}(R_\mathrm{sh})
&=&
\eta {\cal E}\Theta (R_\mathrm{sh}-R_\mathrm{Sedov}) \nonumber\\
&&\times
\left\{ 
\begin{array}{ll}
\frac{1-(R_\mathrm{sh}/R_\mathrm{Sedov})^{-\alpha(2-s)+\beta}}
     {1-(m_{\mathrm{p}}c/p_\mathrm{knee})^{2-s-\beta/\alpha}}    
  & (s+\beta/\alpha \neq 2) \\
& \\
\alpha 
\frac{\log(R_\mathrm{sh}/R_\mathrm{Sedov})}
     {\log(p_\mathrm{knee}/m_{\mathrm{p}}c)} 
  & (s+\beta/\alpha = 2)~~, \\
\end{array} \right.
\label{eq:Qesc2}
\end{eqnarray}
where $\eta$ is the ratio of the total energy of escaping CRs to the explosion energy. 
Note that $\gamma_\mathrm{ad}$ is assumed to be constant. 
Integrating Eq.~(\ref{eq:ush}) with the aid of Eqs.~(\ref{eq:mass}) and (\ref{eq:Qesc2}), the dynamics of the SNR shock is determined.
To make more realistic discussion, we should perform fluid simulations 
with CR back reaction, which is beyond the scope of this paper \citep[e.g.,][]{drury89,voelk08}.

\begin{figure}
\includegraphics[width=20pc]{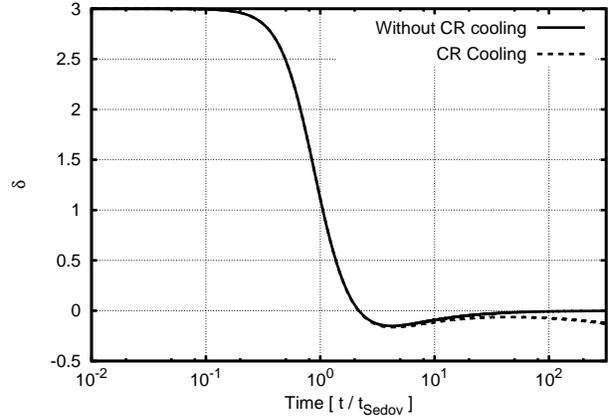}
\caption{ 
The index $\delta=\mathrm{d}\ln E / \mathrm{d}\ln R_\mathrm{sh}$ as a function of time 
for the case PS. 
The solid and dashed lines are for the adiabatic expansion and the 
expansion with CR cooling, respectively.
See the text for details.}
\label{fig1}
\end{figure}
\begin{figure}
\includegraphics[width=20pc]{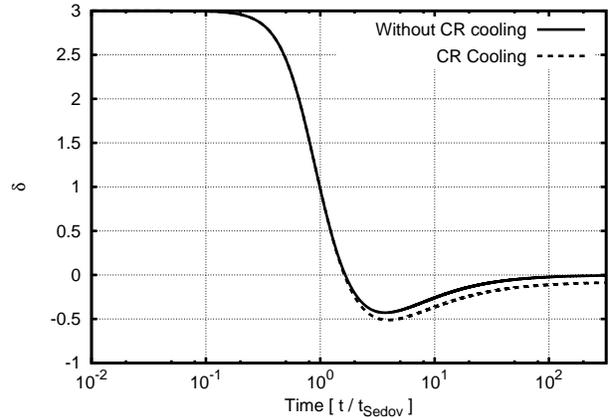}
\caption{The same as Fig.~\ref{fig1}, but for the case PH.}
\label{fig2}
\end{figure}

\subsection{Evolution of $K$}
\label{subsec:evolution_K}

In this subsection, we discuss the evolution of the normalization
factor, $K(R_\mathrm{sh})$, of the spectrum of accelerated particles. 
At present, the injection process for CR acceleration at the shock is not well understood. 
Hence, we consider two representative scenarios of the injection process to model the amount of the accelerated particles.

At first, we consider the same injection model as that of \citet{ptuskin05}.
The model requires that the CR pressure at the shock is proportional to the fluid ram pressure, that is, $P_\mathrm{CR}\propto \rho_\mathrm{am}u_\mathrm{sh}^2$. 
The CR pressure at the shock $P_\mathrm{CR}$ is given by
\begin{equation}
P_\mathrm{CR} \approx \frac{4\pi}{3} 
\int_{m_{\mathrm{p}}c}^{p_{\rm m}(R_\mathrm{sh})} \mathrm{d} pvp^3f_0(p)~~,
\end{equation}
where we neglect the contribution of non-relativistic particles.
Then one can find that $P_\mathrm{CR}$ is related to $f_0(m_{\mathrm{p}}c)$ as
\begin{eqnarray}
P_\mathrm{CR}\propto \left\{ \begin{array}{ll}
f_0(m_{\mathrm{p}}c)    & (s>2, \mathrm{PS}) \\
f_0(m_{\mathrm{p}}c)R_\mathrm{sh}^{\alpha(s-2)} & (s<2, \mathrm{PH}) \\
\end{array} \right. ~~,
\end{eqnarray}
where we have used the fact that the distribution function of CRs at 
the shock front is essentially $f_0(p)\propto p^{-s-2}$ for $p>m_{\mathrm{p}}c$. 
Note that $P_\mathrm{CR}$ is approximately proportional to $f_0(m_{\mathrm{p}}c)$ when $s>2$, because in this case $P_\mathrm{CR}$ only weakly depends on $p_\mathrm{m}$.
Hereafter, the cases of $s>2$ and $s<2$ are called PS and PH, respectively. 
Since $P_\mathrm{CR}\propto \rho_\mathrm{am}u_\mathrm{sh}^2$, the normalization factor of the CR spectrum inside the SNR shock $K(R_\mathrm{sh})$ is calculated as
\begin{eqnarray}
K(R_\mathrm{sh}) \propto R_\mathrm{sh}^3 f_0(m_{\mathrm{p}}c)
\propto \left\{ \begin{array}{ll}
E(R_\mathrm{sh})    & (\mathrm{PS}) \\
E(R_\mathrm{sh}) R_\mathrm{sh}^{\alpha(2-s)} & (\mathrm{PH})~~, \\
\end{array} \right.
\end{eqnarray}
where the mechanical energy of the ejecta is (up to the numerical coefficient)
\begin{equation}
E(R_\mathrm{sh}) \approx \frac{4}{3} \pi \rho_\mathrm{am}u_\mathrm{sh}^2R_\mathrm{sh}^3 \propto u_\mathrm{sh}^2R_\mathrm{sh}^3~~.
\end{equation} 
If the explosion is adiabatic, $E(R_\mathrm{sh})$ is constant with $R_\mathrm{sh}$ during the Sedov phase because $u_\mathrm{sh}\propto R_\mathrm{sh}^{-3/2}$.
However, if the modification of SNR dynamics via CR cooling is taken into account (\S~\ref{sec:SNRshock}), then $u_\mathrm{sh}$ is no longer proportional to $R_\mathrm{sh}^{-3/2}$, and it is obtained by solving Eq.~(\ref{eq:ush}). 
In Figs.~\ref{fig1} and \ref{fig2}, we show $\delta=\mathrm{d} \ln E/\mathrm{d}\ln R_\mathrm{sh}$ as a function of time. 
Figure~\ref{fig1} is for the case  PS in which we adopt $\eta=0.2$, $s=2.23$, $\gamma_\mathrm{ad}=5/3$, and $\beta=-0.2$, while Fig.~\ref{fig2} is for the case PH with parameters $\eta=0.2$, $\gamma_\mathrm{ad}=4/3$,  and $\beta=\alpha(2-s)-0.2$.
One can find that after $t\sim t_\mathrm{Sedov}$, $\delta$ is around $-0.2$, so that $E(R_\mathrm{sh})$ is approximately given by 
$E(R_\mathrm{sh})\propto R_\mathrm{sh}^{\delta}$ with constant 
$\delta\approx-0.2$\footnote{
We search for approximate solutions to
Eqs.~(\ref{eq:ush}), (\ref{eq:Qesc2}) and (\ref{eq:beta})
which determine $\delta$ (and $\beta$) for given parameters
$\alpha$, $s$, $\eta$ and $\gamma_{\rm ad}$.
The procedure is as follows.
For a trial value of $\beta$ which is assumed to be constant,
we solve Eqs.~(\ref{eq:ush}) and (\ref{eq:Qesc2}) to obtain $u_\mathrm{sh}(R_\mathrm{sh})$ so as to calculate $E$ and $\delta$ as functions of $R_\mathrm{sh}$.
It is found that  $\delta$ is almost time-independent after $t_\mathrm{Sedov}$, hence we can derive its average value  in the epoch $t>t_\mathrm{Sedov}$.
Then, we update the value of $\beta$ using Eq.~(\ref{eq:beta}) with
the average value of $\delta$.
We repeat this procedure until the iteration converges.
}.
For the case PS, the effect of CR cooling appears at the late time because low-energy CRs have a large fraction of the total energy of the escaping CRs. 
On the other hand, for the case PH, the effect emerges at an earlier epoch.

Next, we consider  the thermal leakage (TL) model \citep{malkov95}. 
This model requires the continuity of the distribution function $f_0(p)$ to the downstream Maxwelian at the injection momentum $p_{\mathrm{inj}}$, namely
\begin{equation}
f_0(p_{\mathrm{inj}})
\propto \rho_\mathrm{am} /p_{\mathrm{inj}}^3~~.
\end{equation}
Using this fact, we derive
\begin{eqnarray}
K(R_\mathrm{sh}) \propto  R_\mathrm{sh}^3 f_0(m_{\mathrm{p}}c)
\propto \rho_\mathrm{am} R_\mathrm{sh}^3 p_\mathrm{inj}^{s_\mathrm{low}-1} ~~,
\end{eqnarray}
where $s_\mathrm{low}$ is the spectral index in the non-relativistic regime, that is, $f_0(p)\propto p^{-s_\mathrm{low}-2}$ for $p_\mathrm{inj}<p<m_{\mathrm{p}}c$.
We further assume the constant $\rho_\mathrm{am}$ and $p_\mathrm{inj} \propto u_\mathrm{sh}$.
Then we obtain $K(R_\mathrm{sh}) \propto R_\mathrm{sh}^{3+(s_\mathrm{low}-1)(\delta-3)/2}$, so that $\beta=3+(s_\mathrm{low}-1)(\delta-3)/2$. 
Although the value of $s_\mathrm{low}$ is uncertain, one can expect 
$s_\mathrm{low}\geq2$. 
In the test particle approximation, one derives $s_\mathrm{low}=s\geq2$.
If the nonlinear model of shock acceleration is considered,
the spectrum $f_0(p)$ is softer than $p^{-4}$ in the non-relativistic regime \citep{berezhko99}.
In particular, if $s_\mathrm{low}\approx3$, then $\beta\approx\delta$, so that the value of $\delta$ becomes similar to those of PS case.

In summary, the evolution of the normalization factor of the spectrum of accelerated particles is given by $K(R_\mathrm{sh})\propto R_\mathrm{sh}^\beta$, with
\begin{eqnarray}
\beta=\left\{ \begin{array}{ll}
\delta & (\mathrm{PS}) \\
\delta+\alpha (2-s)& (\mathrm{PH}) \\
3 + (s_\mathrm{low}-1)(\delta -3)/2   & (\mathrm{TL}) \\
\end{array} \right. ~~.
\label{eq:beta}
\end{eqnarray}
%

\subsection{The spectrum of escaping particles}

We obtain from Eqs.~(\ref{eq:fesc3}), (\ref{eq:fesc5}), and (\ref{eq:beta}), the index of the momentum spectrum of escaping particles as
\begin{eqnarray}
s_\mathrm{esc}=\left\{ \begin{array}{ll}
s+\delta/\alpha & (\mathrm{PS}) \\
2+\delta/\alpha & (\mathrm{PH}) \\
s+\{3+(s_\mathrm{low}-1)(\delta-3)/2\}/\alpha & (\mathrm{TL}) \\
\end{array} \right. ~~.
\end{eqnarray}
In the following of this section, we discuss which injection model is 
suitable to make the galactic CR spectrum observed at Earth. 
We adopt $\alpha=6.5$ as a typical value (see \S~\ref{sec:SNRpmax}) 
and $s_\mathrm{esc}=2.2$ to make the Galactic CR spectrum. 
In this paper, we assume $s_{\rm esc} \approx 2.2$--2.4, where we believe that the deviation from $2.0$ is significant. 
This would be tested not only from Galactic CR observations but also observations of extragalactic galaxies. 
Note that, even if we adopt $s_\mathrm{esc}=2.4$, 
our results are qualitatively unchanged.

In the case PS, $s_\mathrm{esc}$ is smaller than $s$ because $\delta\la0$, so that the model predicts the harder spectrum of escaping particles than that of the source. However, since $\delta/\alpha\approx0.03$, the difference is small.
In order to reproduce the observed Galactic CR spectrum $s_\mathrm{esc}$, the source spectrum should be $s\approx s_\mathrm{esc}\approx2.2$.
Hence, the PS model requires $s>2$ at the source. 
This condition is satisfied if we consider the diffusive shock acceleration at moderate Mach number \citep{fujita09}.
It is also possible to derive $s>2$ if we consider the effects of neutral particles \citep{ohira09b,ohira09c}.  

In the case PH, because the value of $\delta/\alpha$ is small, $s_\mathrm{esc}$ always near 2, which is the value predicted by the diffusive shock acceleration theory in the strong shock, test-particle limit.
In particular, if $\delta=0$ --- indeed, even in the test-particle limit where the cooling via CR
escape can be neglected, the value of $\delta$ is not exactly zero
unless $t\gg t_\mathrm{Sedov}$ (see Fig.~\ref{fig2}) --- then, one can
find $s_{\rm esc}=2$. This is what \citet{berezhko88} and
\citet{ptuskin05} showed.
Note that from Fig.~\ref{fig2}, $\delta$ is negative for a long time, so that $s_\mathrm{esc}<2$.
Therefore, this model PH cannot reproduce the observed Galactic CR spectrum  at the Earth.

In the case TL, neglecting $\delta$, we obtain 
\begin{equation}
 s_\mathrm{esc}\approx s+\frac{3}{13}(3-s_\mathrm{low})~~. 
\end{equation}
Hence if  $s_\mathrm{low}<3$, then $s_\mathrm{esc}>s$.
For the test particle acceleration, we have $s=s_\mathrm{low}$, so that the source spectrum should be $s\approx2.2$ in order to obtain $s_\mathrm{esc}=2.4$. 
For the nonlinear acceleration, one typically expects $s_\mathrm{low}>2$ and $s<2$, 
so that $s_\mathrm{esc}\leq 29/13 \approx 2.23$. 
Therefore, the nonlinear model seems to explain the inferred source spectrum marginally.
Possible rigorous determination of the value of $s_\mathrm{esc}$ may give a constraint on the theory of nonlinear acceleration.

\section{Application to AGN Cocoon Shocks}
\label{sec:AGN}

In this section, taking account of a constraint derived from the spectrum at the Earth, we study the origin of CRs with energies larger than $\sim10^{17.5}$~eV and the acceleration mechanism of them at AGNs.
There are many works which discuss UHECR production in AGNs 
\citep[e.g.,][]{BS87,Tak90,RB93,berezinsky06,berezhko2008,PMM09}. 
Many of them focus on UHECR acceleration in radio galaxies including 
Fanaroff-Riley (FR) I and II galaxies, which typically have powerful jets. 
In the context of DSA, one can basically suppose three acceleration zones; 
internal shocks in jets, hot spots, and cocoon shocks. 
The former two are most widely discussed scenarios but the detailed study of DSA at such mildly relativistic shocks has not yet been achieved.
In this section, we concentrate on the cocoon shock scenario proposed by \citet{berezhko2008}, where the non-relativistic DSA theory can be applied.

In this scenario, extragalactic CRs with energies larger than the second knee ($\sim10^{17.5}$~eV) may be accelerated at the outer cocoon shock running into the intergalactic medium (IGM). As powerful jets penetrating into a uniform ambient medium with a density $\rho_\mathrm{am}$, the heads of the cocoon advances into the IGM with a velocity $u_\mathrm{h}$.
At the same time, the cocoon expands sideways with a velocity $u_\mathrm{c}$. 
Since the typical cocoon shock is non-relativistic, we apply the 
escape-limited model considered in previous sections.  
Although we hereafter focus on this scenario, note that it is very uncertain whether the efficient acceleration occurs there since observed non-thermal emission is much weaker than that from hot spots and lobes.  

In the following, we investigate whether 
the CR spectrum above the second knee can be explained by 
the AGN cocoon shock scenario with the same 
parameters for young SNRs explaining the CR spectrum below 
the knee, which have been discussed in \S~\ref{sec:snr}.
Similar to the previous calculations in \S~\ref{sec:snr}, we hereafter calculate the values of $\tilde{\alpha}=\mathrm{d}\ln p_m^{(\rm esc)}/\mathrm{d}\ln t$ and $\tilde{\beta}=\mathrm{d}\ln K/\mathrm{d}\ln t$ in order to derive the spectral index, $s_{\rm esc}$.
Here we adopt $\chi=t$, where $\chi$ is a variable appeared in
\S~\ref{sec:spect_escape}.

First, let us consider evolution of the maximum momentum, $p_m^{\rm (esc)}(t)$, in a phenomenological way. 
In the young SNR case (\S~\ref{sec:SNRpmax}), we have phenomenologically expected $p_m \propto t^{-\frac{13}{5}}$. 
Then, by using the Sedov-Taylor solution ($u_{\rm sh} \propto t^{-3/5}$ and $R_{\rm sh}\propto t^{2/5}$), we can easily obtain
\begin{equation}
p_m^{(\rm esc)} (t)  \propto u_{\rm sh}^{\frac{13+2c}{3}} R_{\rm sh}^c ~~,
\label{eq:pm_general}
\end{equation}
where $c$ is a phenomenologically introduced parameter since one can expect $\ell \propto R_{\rm sh}^c$ in general.
Note that $p_m^{\rm (esc)} \propto B \ell u_{\rm sh}$, however, $B$ does not depend on $R_{\rm sh}$ but on $u_{\rm sh}$ as long as $B$ is generated by plasma instabilities such as CR streaming instabilities. 
In the cocoon shock scenario, $u_{\rm sh}$ and $R_{\rm sh}$ are replaced with $u_{\rm c}$ and $R_{\rm c}$, respectively.

In order to obtain the value of $\tilde{\alpha}=\mathrm{d}\ln p_m^{(\rm esc)}/\mathrm{d}\ln t$, the dynamics of the AGN cocoon is necessary. 
A simple consideration of the cocoon dynamics for the constant density IGM tells us that $u_\mathrm{h}$ is almost time-independent and $u_\mathrm{c}$ evolves as $u_{\rm c} \propto t^{-1/2}$ \citep{begelman89}, so that the cocoon radius evolves as $R_\mathrm{c} \propto u_\mathrm{c}t\propto t^{1/2}$ and the jet radius evolves as $R_\mathrm{h} \propto u_\mathrm{h} t\propto t$. 
Then, we obtain $\tilde{\alpha}=\frac{13}{6}$ and $\tilde{\alpha}=2$ for $c=0$ and $c=1$, respectively. 

Next, let us consider the time dependence of the normalization factor of the spectrum of accelerated CRs, $K(t) \propto t^{\tilde{\beta}}$. 
The volume of the acceleration region swept by the cocoon shock is $\sim \mathcal{A} u_\mathrm{c} t$, where $\mathcal{A}$ is the total area of the shock surface. 
If we assume the elliptical shape of the cocoon\footnote{\citet{berezhko2008} assumed a kind of spherical cocoon, i.e.,  $\mathcal{A} \propto R_\mathrm{c} u_\mathrm{c} t$, and used $K(t) \propto \rho_{\rm am} R_{\rm c}^3$ which is similar to the relation obtained in the case of model PS. 
However, when the cocoon becomes spherical, $u_{\rm c}$ and $u_{\rm h}$ evolve according to the adiabatic solution as in the Sedov-Taylor solution for SNRs \citep[][]{begelman89,fujita07}.}, then $\mathcal{A} \propto R_\mathrm{c} u_\mathrm{h}t\propto t^{3/2}$,
so that $\mathcal{A} u_\mathrm{c} t\propto t^{2}$.
In the cases  PS and PH,  $f_0(m_\mathrm{p}c)$ is related to $P_\mathrm{CR}$. 
The dependence of $P_\mathrm{CR}$ is written as 
$P_\mathrm{CR} \propto \rho_\mathrm{am} u_\mathrm{c}^2\propto t^{-1}$, where we neglect for simplicity the evolution of the acceleration efficiency which was considered in \citet{berezhko2008}. 
Then, in the case PS, $f_0(m_\mathrm{p}c) \propto P_\mathrm{CR}$ leads to   
\begin{eqnarray}
K(t) \propto (\mathcal{A} u_\mathrm{c} t) \times f_0(m_\mathrm{p}c) \propto t^{1} ~~,
\end{eqnarray}
while in the case PH, 
$f_0(m_\mathrm{p}c) \propto P_\mathrm{CR} t^{\tilde{\alpha}(2-s)}$ leads to  
\begin{eqnarray}
K(t) \propto (\mathcal{A} u_\mathrm{c} t) \times f_0(m_\mathrm{p}c) \propto t^{1+\tilde{\alpha}(2-s)} ~~.
\end{eqnarray}
Finally,
in the case of model TL (see \S~\ref{subsec:evolution_K}), 
$f_0(m_\mathrm{p}c) \propto \rho_\mathrm{am} p_{\rm inj}^{s_{\rm low}-1}$ leads to  
\begin{eqnarray}
K(t) \propto (\mathcal{A} u_\mathrm{c} t) \times f_0(m_\mathrm{p}c)
  \propto t^{(5-s_{\rm low})/2} ~~,
\end{eqnarray}
where we assume the constant IGM density\footnote{ 
However, it may not be the case since the IGM density may drop with the distance from the nucleus. Then we need to use different adiabatic solutions for $u_\mathrm{c}$ and $R_\mathrm{c}$.}.
Therefore, we obtain
\begin{eqnarray}
\tilde{\beta}=\left\{ \begin{array}{ll}
1    & (\mathrm{PS}) \\
1+\tilde{\alpha}(2-s) & (\mathrm{PH}) \\
(5-s_{\rm low})/2   & (\mathrm{TL}) \\
\end{array} \right. ~~.
\label{eq:beta_agn}
\end{eqnarray}

By using the above results, we can obtain the spectrum of escaping particles. 
Here, we assume $\tilde{\alpha}=2$ expected for $c=1$~($\ell \propto R_c$).
Then from Eq.~(\ref{eq:fesc5}), we derive 
\begin{eqnarray}
s_\mathrm{esc}= \left\{ \begin{array}{ll}
s+1/2 & (\mathrm{PS}) \\
5/2   & (\mathrm{PH}) \\
s+(5-s_{\rm low})/4   & (\mathrm{TL}) \\
\end{array} \right. ~~.
\end{eqnarray}
Interestingly, all the three cases (PS, PH and TL) lead to the source spectral index $s_\mathrm{esc} \approx 2.4$--2.7 which is
 required in the proton dip model and in the other extragalactic scenarios.
In the case of model PS, $s=2.2$ leads to $s_\mathrm{esc} = s + 1/2 = 2.7$. 
In the case of model PH, $s < 2$ leads to $s_{\rm esc} = 2 + 1/2 =2.5$. 
Note that in this case, the value of $s_\mathrm{esc}$ does not depend on $s$ but on $\tilde{\alpha}$ generally. 
Finally in the case of model TL, $s=s_{\rm low}=2$ (based on the test particle acceleration) leads to $s_\mathrm{esc} = 2+3/4=2.75$, which is also consistent with the observed UHECR spectrum in the proton dip model unless the redshift evolution of UHECR sources are fast. If the effect of nonlinear acceleration is prominent, $s < 2$ and $s_{\rm low} > 2$ can give harder indices of $s_{\rm esc} < 2.75$.

\section{Summary and Discussions}
\label{sec:discussion}

In this paper, we have investigated the escape-limited model of CR acceleration,
in which the maximum energy of CRs of an accelerator is limited by the 
escape from the acceleration site.
The typical energy of escaping CRs decreases as the shock decelerates 
because the diffusion length becomes longer. 
After revisiting the escape-limited model and reconsider its detail 
more generally, we have derived a simple relation between the spectrum of 
escaping particles and one in the accelerator. 
Then, using the obtained relation, we have discussed which model of 
injection is potentially suitable to make the Galactic and extragalactic 
CRs observed at the Earth. 

For young SNRs, we have considered the shock evolution with cooling by escaping CRs and those spectra for the three injection models. 
As a result, we have found that in the case PH,
it is difficult to satisfy the condition for the source spectrum of
Galactic CRs ($s_{\rm esc} \approx 2.2$--2.4).
On the other hand, $s_{\rm esc} \approx 2.2$--2.4 can be achieved in cases of PS and TL. 

We have also applied our escape-limited model to AGN cocoon shocks as
well as young SNRs. 
This model is just one of the
various candidates proposed so far, even if AGNs are UHECR
accelerators. 
Nevertheless, it is interesting that the young SNR and the
AGN cocoon shock scenarios can 
explain the Galactic and extragalactic cosmic rays observed
at the Earth in the same picture for all the three injection models
if we accept the proton-dip model inferring 
$s_{\rm esc} \approx2.4$--2.7. 
Whether the proton dip model is real or not can also be tested
by future UHECR and high-energy neutrino observations.    
In this paper, we have focused on the proton case. 
Obviously, heavier nuclei become important above the knee so that we need 
to take into account them in order to explain the CR spectrum over the 
whole energy range. We can also apply the escape-limited model to heavy 
nuclei CRs for this purpose, although it is beyond the scope of this paper.    

We point out a potential problem for 
the magnetic field amplification in the escape-limited model.
In the case of young SNRs, we have determined the evolution of the
maximum energy  in the phenomenological way, and
adopted $\alpha \approx 6.5$. 
Using Eqs.~(\ref{eq:pmax_esc2}) and (\ref{eq:pm_general}),
we obtain $B \propto u_{\rm sh}^{\frac{10+2c}{3}}$, 
where $\ell \propto R_{\rm sh}^c$.
The same result is obtained for  AGN cocoon shocks
because we have considered the case in which the same 
parameters describe both the young SNR shocks and the AGN cocoon shocks.
In particular, for $c =1$ as is used in Eq. (31), we obtain 
$B^2 \propto u_{\rm sh}^8$ which means that $B$ rapidly decreases 
with radius (or time).
In principle, both the dependence of $B$ on $u_{\rm sh}$ and the value of $c$
can  be determined theoretically, and then the evolution of  the maximum
energy should be predicted. 
Some previous works are based on theoretical arguments on the 
magnetic field evolution 
\citep[e.g.,][]{bell04,PLM06,berezhko2008,caprioli09}, 
which seem to be different from our phenomenological one.
At present, the mechanisms of the particle acceleration and
the magnetic field amplification are still highly uncertain
despite of many theoretical efforts  
\citep[e.g.,][]{niemiec08,requelme09, ohira09a, LM09}. 
Hence, we expect that further theoretical and observational studies
might reveal this discrepancy or exclude the possibility of
escape-limited acceleration in the future.

In this paper, we have mainly considered spectra of dispersed CRs around young SNRs and AGN cocoon shocks. 
However, applications to other astrophysical objects are, of course, possible.
For example, the old SNRs detected by {\it Fermi} LAT, such as W28, W44, W51 and IC443 
\citep{abdo09,abdo09w51c} have been of great interest because they likely generate escaping CRs. 
In fact, the number of CRs around such old SNRs is likely to decrease with time or the shock radius, that is $\beta<0$ while $\alpha>0$. 
For example, when we consider the dynamics of an old SNR, we have $u_{\rm sh} \propto t^{-2/3} \propto R_{\rm sh}^{-2}$ \citep[e.g.,][]{yamazaki06}, so that we have $E \propto u_{\rm s}^2R_{\rm s}^3\propto R_{\rm sh}^{-1}$, i.e., $\delta = -1 < 0$.
On the other hand, the value of $\alpha$ may be different from 6.5, which could be attributed to various complications such as the interaction with the dense molecular cloud, and so on.
For example,  for the maximum hardening case, that is, $s_\mathrm{esc}=s
+ (s-1)(\sigma^{-1}-1)/\alpha$ (see Appendix A.2), we find $s_{\rm esc}
\approx 1.5$ when $\alpha \approx 1$ and $s \approx 2.5$ where we assume
$s=(\sigma +2)/(\sigma -1)$.
This might be the case for old SNRs such as W51C \citep{abdo09w51c}.
In addition, the maximum energy may be rather small for the old SNRs, so that the spectrum above $p_m^{(\rm esc)}$ would be suppressed. 
The spectrum of high-energy gamma rays might give us important information on both the acceleration and escape processes of CRs with energies much lower than the knee energy.

\begin{acknowledgements}
We would like to thank Akira~Okumura and Yutaka~Fujita for useful comments.
Y.~O. and K.~M. acknowledge Grant-in-Aid from JSPS.
This work was supported in part 
by grant-in-aid from the 
Ministry of Education, Culture, Sports, Science,
and Technology (MEXT) of Japan,
No.~19047004, No.~21740184, No.~21540259 (R.~Y.).
\end{acknowledgements}

\bibliographystyle{aa}

\appendix
\section{Effect of Adiabatic Loss on the Spectrum of Escaping CRs}
\label{sec:appendix}
The instantaneous spectrum of CRs escaping from the acceleration site 
when the shock radius $R_\mathrm{sh}$ is obtained from the equation (23) of \citet{ptuskin05},
\begin{eqnarray}
\frac{\mathrm{d}f_\mathrm{esc}}{\mathrm{d}R_\mathrm{sh}} &=&\frac{1}{u_\mathrm{sh}}\frac{\mathrm{d}f_\mathrm{esc}}{\mathrm{d}t} \nonumber \\
&=& 4\pi\delta(p-p_\mathrm{m}(R_\mathrm{sh})) \nonumber \\
&&\times \left\{ \frac{\left(1-\frac{1}{\sigma}-\frac{\xi_\mathrm{cr}}{2} \right)}{3}R_\mathrm{sh}^2p_\mathrm{m}f(p_\mathrm{m},R_\mathrm{sh}) \right. \nonumber \\
&&\left. -\int_{0}^{R_\mathrm{sh}}\mathrm{d}rr^2f(p_\mathrm{m},r)\left(\frac{\mathrm{d}p_\mathrm{m}}{\mathrm{d}R_\mathrm{sh}}+\left(1-\frac{1}{\sigma}\right)\frac{p_\mathrm{m}}{R_\mathrm{sh}}\right)  \right\}, \nonumber \\
\label{eq:dfdr}
\end{eqnarray}
where $\sigma$ and $\xi_\mathrm{cr}$ are the total shock compression 
ratio and the ratio of the CR pressure to the shock ram pressure. 
The spectrum and the maximum momentum at the shock are assumed as
\begin{equation}
f(p,R_\mathrm{sh}) = Ap^{-s-2}R_\mathrm{sh}^{\beta-3}~~,
\end{equation}
\begin{equation}
p_\mathrm{m}(R_\mathrm{sh}) = R_\mathrm{sh}^{-\alpha}~~,
\end{equation}
where $A$ is a normalization factor of the distribution function, and
 $p$ and $p_\mathrm{m}$ are normalized by $p_\mathrm{knee}$ while
 $R_\mathrm{sh}$ is normalized by $R_\mathrm{Sedov}$. 
The distribution function of CRs at $r<R_\mathrm{sh}$ can be found by the same method as \citet{ptuskin05}. The fluid velocity at $r<R_\mathrm{sh}$ is
\begin{equation}
u(r,t) = \left(1-\frac{1}{\sigma}\right)\frac{r}{R_\mathrm{sh}(t)}u_\mathrm{sh}(t)~~,
\label{eq:velocity}
\end{equation}
and we solve the following equation
\begin{equation}
\frac{\partial f}{\partial t}+u(r)\frac{\partial f}{\partial r} - 
\frac{1}{3r^2}\frac{\partial}{\partial r}\left(r^2u\left(r\right)\right)p
\frac{\partial f}{\partial p}=0~~.
\end{equation}
Then one can get the following solution. 
\begin{equation}
f(p,r)=Ap^{-s-2}r^{\sigma (s+\beta-1)-s-2}R_\mathrm{sh}^{-(s+\beta-1)(\sigma-1)}~~.
\end{equation}
Therefore the total spectrum of escaping CRs is
\begin{eqnarray}
f_\mathrm{esc}&=&\int \frac{\mathrm{d}f_\mathrm{esc}}{\mathrm{d}R_\mathrm{sh}}\mathrm{d}R_\mathrm{sh} \nonumber \\
&=&f_\mathrm{esc,surface}+f_\mathrm{esc,inside}~~,
\end{eqnarray}
where $f_\mathrm{esc,surface}$ is the first term of Eq.~(\ref{eq:dfdr}) and $f_\mathrm{esc,inside}$ is the second term of Eq.~(\ref{eq:dfdr}). These are given by
\begin{equation}
f_\mathrm{esc,surface}=\frac{4\pi A}{\alpha}\frac{1-\frac{1}{\sigma}-\frac{\xi_\mathrm{cr}}{2}}{3}p^{-s-2-\beta/\alpha},
\end{equation}
\begin{eqnarray}
f_\mathrm{esc,inside}&=&4\pi A\left(\alpha-1+\frac{1}{\sigma}\right) \nonumber \\
&&\times \int_{0}^{R_\mathrm{sh}}\mathrm{d}R \delta(p-p_\mathrm{m}(R))p_\mathrm{m}^{-1-s}R^{-(s+\beta-1)(\sigma-1)-1} \nonumber \\
&&\times \int_{0}^{R}\mathrm{d}rr^{\sigma(s+\beta-1)-s}~~.
\label{eq:finside}
\end{eqnarray}
The $\mathrm{d}r$ integral of Eq.~(\ref{eq:finside}) is approximately given by
\begin{eqnarray}
&&\int_{0}^{R}\mathrm{d}rr^{\sigma(s+\beta-1)-s} \nonumber \\
&&~~~~\propto \left\{ \begin{array}{ll}
R^{\sigma(s+\beta-1)-s+1} & \left[\beta>\left(s-1\right)\left(\frac{1}{\sigma}-1\right)\right] \\
R_0^{\sigma(s+\beta_\mathrm{free}-1)-s+1} & \left[\beta<\left(s-1\right)\left(\frac{1}{\sigma}-1\right)\right]~~, \\
\end{array} \right.
\label{eq:inside}
\end{eqnarray}
where $R_0$ is the radius at which $\beta_\mathrm{free}$ ($> \left(s-1\right)\left(\sigma^{-1}-1\right)$) changes to $\beta$ ($< \left(s-1\right)\left(\sigma^{-1}-1\right)$). Here, we assume that $\beta$ changed when $R_\mathrm{ sh}=1$, that is, $t=t_\mathrm{sedov}$. Then from Eq.~(\ref{eq:velocity}), $R_0 = R_\mathrm{sh}^{1-\sigma^{-1}}$.

\subsection{$\beta>\left(s-1\right)\left(\frac{1}{\sigma}-1\right)$}
In this case, the $\mathrm{d}r$ integral of Eq.~(\ref{eq:finside}) is 
dominated by the outer region, that is, spectrum of escaping particles 
does not depend on the past acceleration history. $f_\mathrm{esc,inside}$ is 
calculated as
\begin{eqnarray}
f_\mathrm{esc,inside} &=& \frac{4\pi A(\alpha-1+\frac{1}{\sigma})}{1-s+\sigma (s+\beta-1)} \nonumber \\
&&\times \int_{0}^{R_\mathrm{sh}}\mathrm{d}R \delta(p-p_\mathrm{m}(R))p_\mathrm{m}^{-1-s}R^{\beta-1} \nonumber \\
&=& \frac{4\pi A}{\alpha}\frac{\alpha-1+\frac{1}{\sigma}}{1-s+\sigma (s+\beta-1)}p^{-s-2-\beta/\alpha}.
\end{eqnarray}
Hence, as long as $\beta>\left(s-1\right)\left(\sigma^{-1}-1\right)$, 
the energy spectrum index of escaping CRs  is $s_\mathrm{esc}=s+\beta/\alpha$.

\subsection{$\beta<\left(s-1\right)\left(\frac{1}{\sigma}-1\right)$}
In this case, the $\mathrm{d}r$ integral of Eq.~(\ref{eq:finside}) is 
dominated by the inner region, that is, the spectrum of escaping particles depends on the past acceleration history. Especially, only the acceleration at $t=t_\mathrm{sedov}$ is important, and then $f_\mathrm{esc,inside}$ is calculated as
\begin{eqnarray}
f_\mathrm{esc,inside} &=& \frac{4\pi A(\alpha-1+\frac{1}{\sigma})}{1-s+\sigma (s+\beta_\mathrm{free}-1)} \nonumber \\
&&\times \int_{R_0}^{R_\mathrm{sh}}\mathrm{d}R \delta(p-p_\mathrm{m}(R))p_\mathrm{m}^{-1-s}R^{(s-1)(\frac{1}{\sigma}-1)-1} \nonumber \\
&=& \frac{4\pi A}{\alpha}\frac{\alpha-1+\frac{1}{\sigma}}{1-s+\sigma (s+\beta_\mathrm{free}-1)}p^{-s-2-(s-1)(\frac{1}{\sigma}-1)/\alpha}. \nonumber \\
\end{eqnarray}
Because $f_\mathrm{esc,inside}$ is softer than $f_\mathrm{esc,surface}$ (at $p < p_{\rm m}$), $f_\mathrm{esc,inside}$ is larger than $f_\mathrm{esc,surface}$.
Hence the spectral index of escaping CRs is 
$s_\mathrm{esc}=s+(s-1)(\sigma^{-1}-1)/\alpha$ and does not depend on $\beta$. 
The largest possible hardening from the spectral index of the acceleration site $\Delta s = s-s_\mathrm{esc}$ is
\begin{eqnarray}
\Delta s &=&\frac{s-1}{\alpha}\left(1-\frac{1}{\sigma}\right) \nonumber\\
&=&\frac{3(s-1)}{\alpha(s+2)}~~,
\end{eqnarray}
where we assume the relation between the spectrum index and the compression ratio, $s=(\sigma+2)/(\sigma-1)$.

\end{document}